\pdfoutput=1

\PassOptionsToPackage{table}{xcolor}
\documentclass[11pt]{article}

\usepackage[preprint]{acl}

\usepackage{times}
\usepackage{latexsym}

\usepackage[T1]{fontenc}

\usepackage[utf8]{inputenc}

\usepackage{microtype}

\usepackage{inconsolata}

\usepackage{graphicx}

\title{Scientific Paper Retrieval with LLM-Guided Semantic-Based Ranking}

\author{Yunyi Zhang\textsuperscript{1},
Ruozhen Yang\textsuperscript{1},
Siqi Jiao\textsuperscript{1},
SeongKu Kang\textsuperscript{2},
Jiawei Han\textsuperscript{1}\\
\textsuperscript{1}University of Illinois Urbana-Champaign
\textsuperscript{2}Korea University\\
\texttt{\{yzhan238,ruozhen2,sjiao2,hanj\}@illinois.edu\quad seongkukang@korea.ac.kr}
}

\usepackage{amsmath, amsfonts, amsthm, xspace, color}
\usepackage{booktabs} 

\usepackage{multirow, tabularx} 
\usepackage{booktabs} 
\usepackage{enumitem}
\usepackage[T1]{fontenc}
\usepackage{epstopdf}
\usepackage{graphicx}
\usepackage{url}
\usepackage{wrapfig,lipsum}
\usepackage{makecell}
\usepackage{wrapfig}
\usepackage{latexsym}

\usepackage{microtype}

\usepackage{caption}
\usepackage{subcaption}

\definecolor{darkgreen}{rgb}{0.0, 0.5, 0.0}

\newcommand{\hide}[1]{} 

\newcommand{\ie}{i.e.\xspace} 
\newcommand{\eg}{e.g.\xspace} 
\newcommand{\nop}[1]{}

\newtheoremstyle{exampstyle}
  {0pt} 
  {0pt} 
  {\itshape} 
  {1em} 
  {\bfseries} 
  {.} 
  {.5em} 
  {} 

\theoremstyle{exampstyle}
\newtheorem{thm:def}{Definition}

\newtheorem{thm:eg}{Example}
\newtheorem{thm:lem}{Lemma}
\newtheorem{thm:obs}{Observation}
\newtheorem{thm:req}{Requirement}
\newtheorem{thm:prop}{Proposition}
\newtheorem{thm:principle}{Principle}
\newtheorem{thm:thm}{Theorem}
\newtheorem{thm:corollary}{Corollary}

\def \L {\mathcal{L}}

\def \D {\mathcal{D}}

\def \T {\mathcal{T}}

\newcommand{\Our}{\mbox{SemRank}\xspace}

\usepackage[export]{adjustbox}
\usepackage[super]{nth}
\usepackage{dsfont}

\usepackage[noend,ruled,linesnumbered]{algorithm2e}
\usepackage[most]{tcolorbox}
\usepackage{soul}
\sethlcolor{pink}

\begin{document}
\maketitle
\begin{abstract}
Scientific paper retrieval is essential for supporting literature discovery and research. While dense retrieval methods demonstrate effectiveness in general-purpose tasks, they often fail to capture fine-grained scientific concepts that are essential for accurate understanding of scientific queries. Recent studies also use large language models (LLMs) for query understanding; however, these methods often lack grounding in corpus-specific knowledge and may generate unreliable or unfaithful content. To overcome these limitations, we propose \Our, an effective and efficient paper retrieval framework that combines LLM-guided query understanding with a concept-based semantic index. Each paper is indexed offline using multi-granular scientific concepts, including general research topics and detailed key phrases. At query time, an LLM identifies core concepts derived from the corpus to explicitly capture the query's information need. These identified concepts enable precise semantic matching, significantly enhancing retrieval accuracy.
Experiments show that \Our consistently improves the performance of various base retrievers, surpasses strong LLM-based baselines, and remains highly efficient.\footnote{Code can be found at: \url{https://github.com/yzhan238/SemRank}.}
\end{abstract}


\section{Introduction}

Scientific paper retrieval is a crucial task to facilitate literature discovery and accelerate scientific progress~\cite{TaxoIndex}. Unlike general-purpose information retrieval, scientific paper retrieval is more challenging because queries often involve theme-specific intent and specialized terminology. In addition, acquiring labeled query-passage pairs for supervised fine-tuning is costly and requires domain expertise, making it impractical to continuously annotate more data to adapt the fast-evolving scientific domains.

Recently, dense passage retrieval methods have been widely studied in various ad-hoc searches~\cite{DPR,Izacard2021UnsupervisedDI}. These methods encode the overall semantics of queries and passages into the same vector space and measure relevance using vector similarity. Although being effective in different general domain applications, they still face challenges in scientific paper retrieval. 

Specifically, general-purpose semantic representations learned by dense retrievers often fail to capture fine-grained scientific concepts that are crucial for accurately understanding and satisfying a scientific query~\cite{chen-etal-2022-salient, shavarani-sarkar-2025-entity}. For example, the query ``\textit{Can you point me to studies discussing methods for evaluating text generation models on various dimensions?}'' involves not only general topics like ``natural language generation'' and ``automatic evaluation'' that need to be inferred from the text, but also specific details like ``multidimensional evaluation''. A dense retriever, however, only encodes the text in a holistic view, while it lacks the ability and controllability to focus on the scientific concepts which are the core need of the query.

\begin{figure*}[t]
  \centering
  \includegraphics[width=0.99\textwidth]{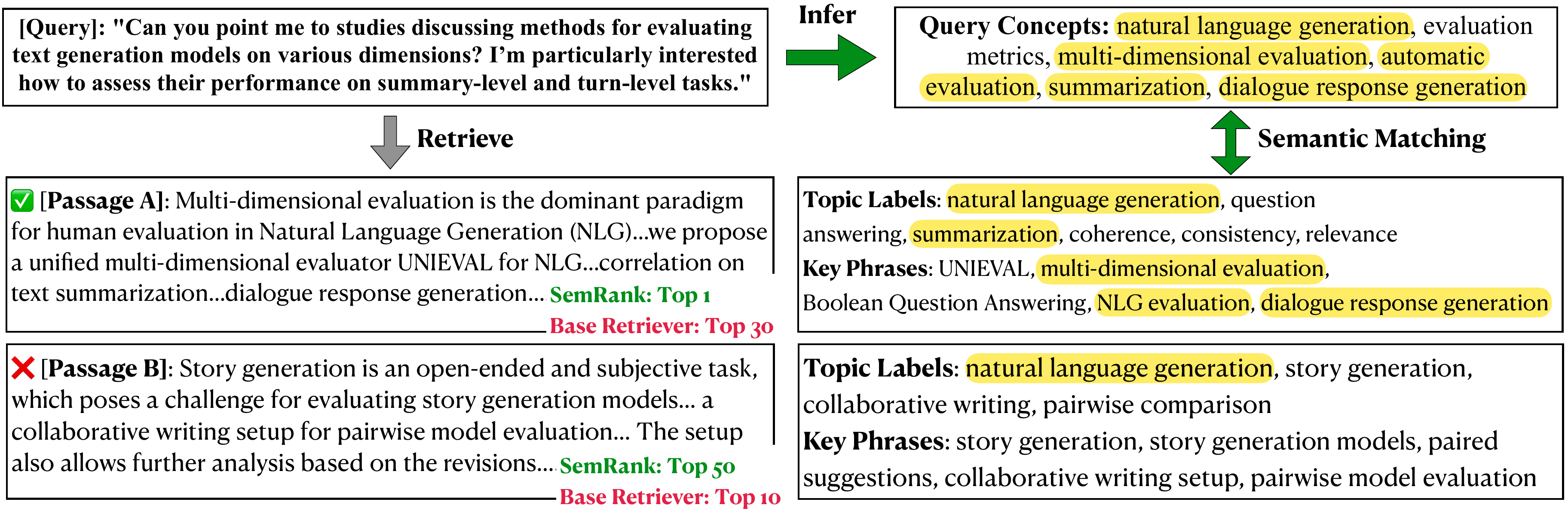}
  \vspace*{-0.25em}
  \caption{An illustrative example from LitSearch with SPECTER-v2 as base retriever. By capturing the scientific concepts for corpus and query, \Our substantially improves the ranking results.}
  \label{fig:intro-example}
  \vspace*{-1em}
  \label{fig:example}
\end{figure*}

With the advancements of large language models (LLM) such as GPT~\cite{OpenAI2023GPT4TR} and Claude~\cite{claude3}, recent studies also explore how to utilize LLMs in query understanding to help retrieval tasks. For example, HyDE~\cite{gao-etal-2023-precise} uses an LLM to generate a hypothetical passage for encoding, and CSQE~\cite{lei-etal-2024-corpus} prompts an LLM to select a set of relevant sentences to expand the original query. However, these methods still rely on a pre-trained dense retriever to encode overall semantics on the document or sentence levels, lacking the ability to explicitly capture what the query is asking for. In addition, LLMs are not inherently retrieval models. They do not have the vast and dynamic knowledge in the scientific literature necessary to understand scientific queries. Therefore, when used in a zero-shot manner, LLMs cannot identify domain-specific terminology and may generate hallucinated content. Thus, how to effectively augment LLMs with corpus-based knowledge while capturing specific query information remains a challenge.

To overcome these limitations, we propose to utilize LLMs' text understanding ability in the scientific retrieval task with the help of a scientific concept-based semantic index. Unlike earlier studies which construct domain-specific semantic index at topic level~\cite{BioASQ}, we build concept-based semantic index at various granularities: from broad research topics such as ``natural language generation'' to specific terms such as ``multidimensional evaluation metrics''. These multi-granular concepts capture the essential content of the paper. Then, we use the semantic index to improve any existing retrievers with the help of LLMs. Specifically, we prompt an LLM to identify a set of core concepts to explicitly represent what the scientific query is asking for. We augment the prompt with candidate concepts derived from the corpus, which helps the LLM to reduce hallucination and ensure the generated content align with the semantic index of corpus. Figure~\ref{fig:intro-example} shows an example that, by accurately capturing the multi-granular concepts of the corpus and query, we can improve the scientific paper retrieval results by focusing on the core need of the query.

We propose the \Our, LLM-Guided \textbf{Sem}antic-Based \textbf{Rank}ing, a plug-and-play framework for scientific paper retrieval. First, during the indexing time, we build scientific concept-based semantic index for the corpus by identifying a set of research topics and key phrases for each paper. To ensure the topic labels in their canonical forms, we train an auxiliary topic classifier model to identify a set of candidate topics from a large label space\footnote{An academic label space is widely available such as \href{https://dl.acm.org/ccs}{ACM CCS} and Microsoft Academic Graph~\cite{sinha2015overview}}. Then, we prompt an LLM to select the core topic labels and extract key phrases to build the semantic index.
Then given a query during retrieval, we first construct a set of candidate concepts from the corpus using a base retriever, from which an LLM can be augmented with corpus-based knowledge and identify a set of core concepts for the query. These concepts, serving as an explicit guidance on what the query is asking for, will be used for concept-level semantic matching to improve the retrieval results.

\begin{figure*}[!t]
    \centering
    \centerline{\includegraphics[width=0.99\textwidth]{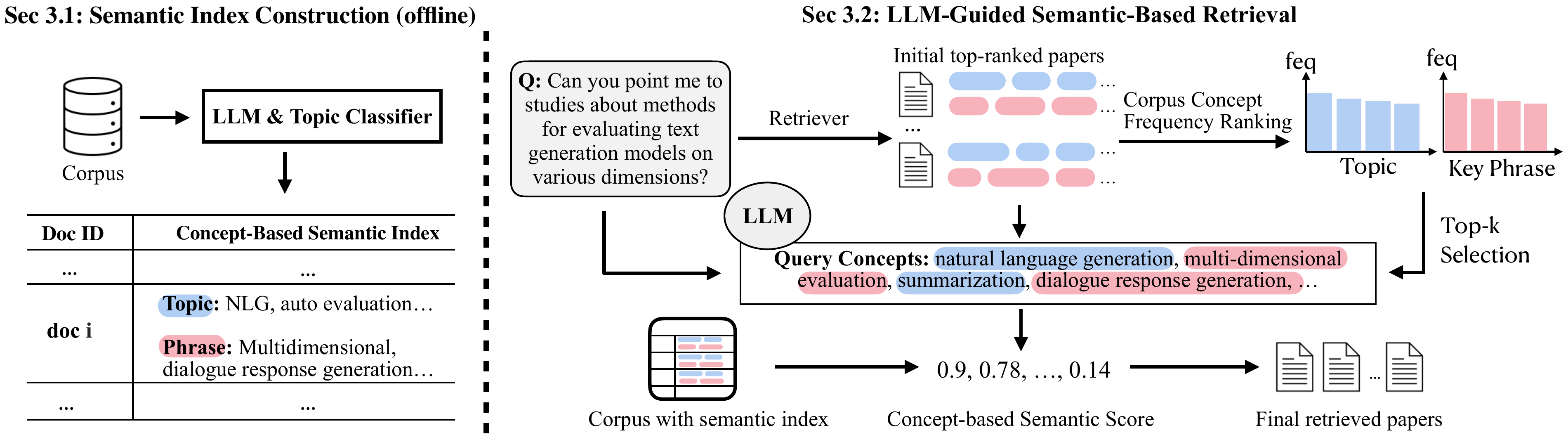}}
    \vspace*{-0.25em}
    \caption{Overview of the \Our framework.}
    \label{fig:framework}
    \vspace*{-1em}
\end{figure*}

Our method can be easily integrated with any dense retriever and improve their retrieval quality without relying on any annotated query data.
Experiments show that our method is both effective and efficient. During retrieval time, \Our only needs one LLM prompting per query, no additional call to the base retriever, and all computation can be done easily on CPUs. Yet it can significantly improve the ranking performance of a wide range of base retrievers and outperforms various LLM-based baselines.

The contributions of this paper are as follows:
\begin{itemize}[leftmargin=*,nosep]
    \item We propose \Our, a plug-and-play framework for scientific paper retrieval, which utilizes concept-based semantic index and LLM guidance to explicit capture the information need for scientific queries.
    \item We develop a light-weighted method which augments an LLM with semantic index of corpus to accurately identify a set of core scientific concepts for the query, which are then used to improve the retrieval performance through concept-level semantic matching.
    \item Through extensive experiments, we show that \Our consistently improves the base retrievers' performance and outperforms existing baselines while being more efficient.
\end{itemize}
\section{Problem Formulation}

Given a corpus $\D$ of scientific papers, our goal is to retrieve and rank the papers according to their relevance to a given query $q$. Specifically, we will build concept-based semantic index for the corpus by identifying a set of scientific concepts representing the core information of each paper. We assume a scientific topic label space $\T$, which is widely available and normally contains a large number of research topics. In this work, we use Microsoft Academic Graph~\cite{sinha2015overview}, which contains 13,613 Computer Science topics. Then, we utilize LLMs and the semantic index to improve the scientific paper retrieval performance.
\section{Methodology} \label{sec:method}

In this section, we will present our \Our framework. We first introduce the offline semantic index construction module (Sect. \ref{sec:indexing}), then we introduce the LLM-guided semantic-based ranking for scientific paper retrieval (Sect. \ref{sec:retrieval}). Figure~\ref{fig:framework} shows an overview of \Our.

\subsection{Semantic Index Construction} \label{sec:indexing}

To capture the core research concepts at various granularities for scientific papers, we propose to build a semantic index of the corpus that contains general research topics and specific key phrases. The broad topics aim to cover the overall themes of the papers that are not explicitly mentioned such as ``natural language generation'' and ``automatic evaluation,'' while the key phrases aim to capture the detailed information specific to the paper such as ``multidimensional valuation metrics''. Such a well-structured corpus foster flexible query matching at different granularities. 

While directly prompting LLMs could be a viable solution to assign each document a set of topics and key phrases, it is hard to ensure the faithfulness of the LLM-generated content. Besides, the research topics are often not explicitly mentioned in the text and LLMs can generate topics at random granularity, which is less controllable and hard to match during retrieval. Therefore, we propose to first fine-tune a multi-label topic classifier for scientific papers with a domain-specific topic label space, from which we can obtain a set of candidate topics for each scientific paper.

\paragraph{Candidate Topic Prediction} 
We first fine-tune a multi-label text classifier to estimate the likelihood of a paper belonging to a research topic. We use a simple log-bilinear text matching network as our model architecture~\cite{zhang2025teleclass}. We initialize the paper encoder with a pre-trained scientific domain language model (e.g., SPECTER-v2~\cite{specter}). We also get the topic embeddings using the same pre-trained model and detach them from the encoder. This ensures only the embeddings are updated without back-propagating to the backbone for saving cost. Then, the classifier predicts the probability of document $d_i$ belonging to topic $t_j \in \T$ with log-bilinear matching: 
\begin{equation*}\label{eq:log-bilinear}
    p(t_j|d_i) = \sigma(\exp{(\mathbf{t}_j^T\mathbf{W}\mathbf{d}_i)}),
\end{equation*}
where $\sigma$ is the sigmoid function, $\mathbf{W}$ is a learnable interaction matrix, and $\mathbf{t}_j$ and $\mathbf{d}_i$ are the encoded topic and document.

We fine-tune the multi-label topic classifier using the binary cross entropy loss. Given the positive topics of paper $d_i \in \D$ in the label space $\T_i \subset \T$, we train the classifier with:
\begin{align*}
    \L = -\sum_{d_i \in \D} &\Big( \sum_{t_j \in \T_i} \log{p(t_j|d_i)}\\ 
    &+ \alpha\sum_{t_j \notin \T_i} \log{(1-p(t_j|d_i))}\Big),
\end{align*}
where the coefficient $\alpha$ is a small constant number to counter the imbalanced numbers of positive and negative labels in a large label space.

\paragraph{Core Concepts Identification}

With the fine-tuned scientific topic classifier, we first predict a set of candidate topics likely relevant to each paper in the retrieval corpus. Then, we prompt an LLM to perform two tasks: (1) select a set of topics from the candidate list that are not too broad or irrelevant, and (2) extract a set of key phrases from the paper. By providing a candidate topic list, we turn a generation task (of zero-shot prompting) to a selection and extraction task for LLMs, which reduces the chance of hallucination and also ensures the selected topics following a label space. Figure~\ref{fig:prompt-sem-idx} shows our prompt.

By first predicting candidate topic labels with a text classifier, not only can we resolve the issue of LLMs' poor performance on a large structured label space~\cite{yu-etal-2023-instances, zhang2025teleclass}, but also reduce the cost of prompting LLMs by 98\%. A more detailed cost analysis is in Section~\ref{sec:efficiency}. Also note that we only need to train one topic classifier for a domain and use it for all corpora in the same domain (\eg, Computer Science).

In summary, for each scientific paper $d_i \in \D$, we identify a set of research topics belonging to a label space, denoted as $T_i \in \T$, and a set of key phrases extracted from it, denoted as $P_i$.

\subsection{LLM-Guided Semantic-Based Retrieval} \label{sec:retrieval}

With the constructed semantic index capturing the core concepts discussed in the corpus, we now present how \Our leverages it to enhance retrieval by explicitly modeling a query's information need. Given a base retriever, \Our first identifies a set of candidate concepts relevant to the query and prompts an LLM to analyze the retrieval context and select the most salient core concepts. These core concepts are then used to refine the initial retrieval results, yielding a concept-aware ranking that better aligns with the query intent.

\paragraph{Candidate Concepts Construction for Query}
Directly prompting an LLM to assign a set of scientific concepts to a query is not optimal, because the generated content may not align with the semantic index structure of the corpus and thus make it difficult to match between the query and the papers. Therefore, we propose to first construct a set of candidate concepts for a query from the corpus.

Inspired by the idea of pseudo relevance feedback, given a base retriever $s^{base}$ and its initial ranking results, we collect a set of topics and key phrases that are frequently mentioned in the top-ranked papers. Specifically, for a query $q$, we collect lists of top-k most frequent topics and key phrases mentioned by the top-ranked papers using their pre-constructed semantic index $T_i$ and $P_i$, and denote the lists as $T^0(q)$ and $P^0(q)$. These concepts, being frequently mentioned by top-ranked papers, are likely relevant to the initial query.

\paragraph{LLM-Guided Core Concept Identification}
With the constructed candidate concepts of the query, we can now prompt an LLM to identify a set of core concepts for the query that can most likely identify its relevant papers. Given the topic and key phrases lists $T^0(q)$ and $P^0(q)$ as well as the top-k papers in the current ranking $D^0(q)$, we instruct an LLM to select a set of concepts from the candidate lists in order to improve the current ranking results. Figure~\ref{fig:prompt-core-concept} shows our prompt.

Because the candidate concepts are collected from the semantic index, they contribute to two advantages: (1) they serve as a high-level summary of the current retrieval results to help the LLM interpret the query and current results; (2) they provide a high-quality candidate list for the LLM, ensuring the faithfulness and reducing hallucination.

After prompting the LLM, we get a set of core concepts selected from the candidate lists, which we denote as $C(q)$. This set contains the most important concepts relevant to the original query and thus explicitly represents the information need of the query. Also, the concepts are in different granularities, proper for matching at the needed level of details. For example, for the query shown in Fig~\ref{fig:intro-example}, the LLM identifies general topics like ``natural language generation'' and ``automatic evaluation'', and specific terms like ``multidimensional evaluation'' and ``dialogue response generation''.

\paragraph{Core Concept-Based Ranking} 
Finally, with the identified set of core concepts of the query, we can use the semantic index of the corpus to re-evaluate the ranking. Specifically, given the core concepts of the query $C(q)$ and the the concepts of a paper $C_i = T_i \cup P_i$, we calculate their similarity with a multi-vector similarity matching score.
\begin{equation*}
    s^{sem}(q, d_i) = \frac{1}{|C(q)|} \sum_{c \in C(q)} \max_{c' \in C_i} sim(\mathbf{c}, \mathbf{c'}),
\end{equation*}
where $\mathbf{c}$ denotes the embedding of a concept by a semantic encoder (e.g., SPECTER-v2) and $sim()$ represents the cosine similarity function. This embedding-based soft matching process identifies the most similar concept in a paper for each query's concept, which accounts for the situation where similar concepts are expressed slightly differently (\eg, ``hallucination'' and ``hallucinated content''). 
Then, we combine this semantic-based score with the base retriever's score by z-score normalization (denoted by $z(\cdot)$), based on which we can re-rank the papers to get a new ranked list.
\begin{equation*}
    s(q, d_i) = z(s^{base}(q, d_i)) + z(s^{sem}(q, d_i)).
\end{equation*}

\paragraph{Efficiency of \Our Retrieval}
Given the ranking results of a base retriever, our LLM-guided semantic-based ranking process is highly efficient. First, it only requires one LLM call, and the output length is minimal because it is highly-structured with a list of scientific concepts. Second, because all the query's concepts are selected from the semantic index and all concept embeddings can be pre-computed offline, we only need the cosine similarity computation (i.e., dot product with normalized vectors) during retrieval, which is highly efficient and can be done on CPUs. Besides, the pairwise similarities between concepts can also be pre-computed for the maximal inference efficiency, but it requires substantial amount of storage, so we opt to compute the similarity during retrieval. We conduct a detailed efficiency analysis in Sect.~\ref{sec:efficiency}.
\section{Experiments}

\subsection{Experiment Setup} \label{sec:exp-setup}

\begin{table}[t!]
\centering
\caption{Datasets overview.}\label{table:dateset}
\vspace*{-0.5em}
\scalebox{0.8}{
    \begin{tabular}{cccc}
        \toprule
        \bf Dataset & \bf corpus size & \bf \# test query & \bf doc/query\\
        \midrule
        \bf CSFCube & 4,207 & 34 & 13.32 \\
        \bf DORISMAE & 8,482 & 90 & 19.49 \\
        \bf LitSearch & 64,183 & 597 & 1.07 \\
        \bottomrule
    \end{tabular}
}
\vspace*{-1em}
\end{table}

\paragraph{Datasets}
We use three public datasets on scientific paper retrieval: \textbf{CSFCube}~\cite{mysore2021csfcube}, \textbf{DORISMAE}~\cite{dorismase}, and \textbf{LitSearch}~\cite{ajith-etal-2024-litsearch}, including both human-annotated and LLM-generated relevance labels. We use the processed version of CSFCube and DORISMAE released by \citet{TaxoIndex}\footnote{https://aclanthology.org/attachments/2024.emnlp-main.407.data.zip} and LitSearch from its official github\footnote{https://github.com/princeton-nlp/LitSearch}. Table~\ref{table:dateset} summarizes the overall statistics of the datasets.

\begin{table*}[!t]
\centering
\caption{Performance of baselines in Recall@K, with the best score \textbf{boldfaced} and the second best \underline{underlined}.}
\label{table:baseline-res}
\vspace*{-0.5em}
\scalebox{0.9}{
\begin{tabular}{cl|cc|cc|ccc}
\toprule
\multirow{2}{*}{\bf Base} & 
\multirow{2}{*}{\bf Methods}    & \multicolumn{2}{c}{\bf CSFCube}                  & \multicolumn{2}{|c}{\bf DORISMAE}   & \multicolumn{3}{|c}{\bf LitSearch}                   \\
\cmidrule{3-4} \cmidrule{5-6} \cmidrule{7-9}
\multicolumn{2}{c|}{}                            & R@50    & R@100    & R@50    & R@100    & R@5    & R@20    & R@100 \\
\midrule
\parbox[t]{2mm}{\multirow{7}{*}{\rotatebox[origin=c]{90}{SPECTER-v2}}}
& Retriever & 0.5331 & 0.6860 & 0.5305 & 0.7208 & 0.3931 & 0.5551 & 0.7205\\
& Boudin et al. & 0.5088 & 0.6739 & 0.4695 & 0.6040 & 0.4345 & 0.5671 & 0.7250 \\
& BERT-QE &0.5243 & 0.6689 & 0.5284 & 0.6942 & 0.3959 & 0.5551 & 0.7364\\
& ToTER &0.5443 & 0.7131 & 0.5319 & 0.7234 & 0.3948 & 0.5568 & 0.7239\\
& HyDE & \underline{0.5879} & \underline{0.7473} & 0.5110 & 0.6789 & 0.4241 & \underline{0.5923} & \underline{0.7682}\\
& GRF & 0.5599 & 0.6758 & \underline{0.5442} & \underline{0.7283} & 0.4319 & 0.5830 & 0.7604\\
& CSQE & 0.5586 & 0.7149 & 0.5022 & 0.6491 & \underline{0.4366} & 0.5747 & 0.7223\\
& \cellcolor{gray!15}\Our & \cellcolor{gray!15}\bf 0.6222 & \cellcolor{gray!15}\bf 0.7601 & \cellcolor{gray!15}\bf 0.5894 & \cellcolor{gray!15}\bf 0.7451 & \cellcolor{gray!15}\bf 0.5028 & \cellcolor{gray!15}\bf 0.6316 & \cellcolor{gray!15}\bf 0.7746\\
\midrule
\parbox[t]{2mm}{\multirow{7}{*}{\rotatebox[origin=c]{90}{E5-large-v2}}}
& Retriever & 0.6111 & 0.7362 & 0.5548 & 0.7162 & 0.5137 & 0.6573 & 0.7765\\
& BERT-QE &0.6399 & 0.7589 & 0.5943 & 0.7451 & 0.4906 & 0.6361 & 0.7881\\
& ToTER &0.6134 & 0.7553 & 0.5596 & 0.7120 & 0.4981 & 0.6627 & 0.7951\\
& HyDE & 0.6203 & 0.7255 & 0.5559 & 0.7236 & 0.4854 & 0.6592 & 0.8149\\
& GRF & 0.6183 & \underline{0.7797} & \underline{0.6025} & \underline{0.7501} & 0.5347 & \underline{0.6984} & \bf 0.8319\\
& CSQE & \underline{0.6416} & 0.7549 & 0.4787 & 0.5977 & \underline{0.5503} & 0.6389 & 0.7719\\
& \cellcolor{gray!15}\Our & \cellcolor{gray!15}\bf 0.6661 & \cellcolor{gray!15}\bf 0.8177 & \cellcolor{gray!15}\bf 0.6286 & \cellcolor{gray!15}\bf 0.7754 & \cellcolor{gray!15}\bf 0.5807 & \cellcolor{gray!15}\bf 0.7042 & \cellcolor{gray!15}\underline{0.8312}\\
\bottomrule
\end{tabular}
}
\vspace*{-0.5em}
\end{table*}

\begin{table*}[!t]
\centering
\caption{Performance of \Our on three datasets with different base retrievers, with the best score \textbf{boldfaced}.}
\label{table:base-retriever-res}
\vspace*{-0.5em}
\scalebox{0.9}{
\begin{tabular}{l|cc|cc|ccc}
\toprule
\multirow{2}{*}{\bf Methods}    & \multicolumn{2}{c}{\bf CSFCube}                  & \multicolumn{2}{|c}{\bf DORISMAE}   & \multicolumn{3}{|c}{\bf LitSearch}                   \\
\cmidrule{2-3} \cmidrule{4-5} \cmidrule{6-8}
                            & R@50    & R@100    & R@50    & R@100    & R@5    & R@20    & R@100 \\
\midrule
SPECTER-v2 & 0.5331 & 0.6860 & 0.5305 & 0.7208 & 0.3931 & 0.5551 & 0.7205\\
\quad+ \Our & \bf 0.6222 & \bf 0.7601 & \bf 0.5894 & \bf 0.7451 & \bf 0.5028 & \bf 0.6316 & \bf 0.7746\\
\midrule
BM25 & 0.4651 & 0.5966 & 0.5721 & 0.7441 & 0.4381 & 0.5794 & 0.7362\\
\quad+ \Our & \bf 0.5840 & \bf 0.7076 & \bf 0.6507 & \bf 0.8104 & \bf 0.5126 & \bf 0.6612 & \bf 0.7920\\
\midrule
Hybrid & 0.5855 & 0.7131 & 0.6743 & 0.8297 & 0.5397 & 0.6877 & 0.7881\\
\quad+ \Our & \bf 0.6434 & \bf 0.7673 & \bf 0.6779 & \bf 0.8358 & \bf 0.5835 & \bf 0.7161 & \bf 0.7922\\
\midrule
E5-large-v2 & 0.6111 & 0.7362 & 0.5548 & 0.7162 & 0.5137 & 0.6573 & 0.7765\\
\quad+ \Our & \bf 0.6661 & \bf 0.8177 & \bf 0.6286 & \bf 0.7754 & \bf 0.5807 & \bf 0.7042 & \bf 0.8312\\
\midrule
GritLM-7B & 0.6732 & 0.7742 & 0.6415 & 0.8037 & 0.6908 & 0.8001 & 0.9149\\
\quad+ \Our & \bf 0.7290 & \bf 0.8466 & \bf 0.6743 & \bf 0.8187 & \bf 0.6955 & \bf 0.8171 & \bf 0.9221\\
\bottomrule
\end{tabular}
}
\vspace*{-0.5em}
\end{table*}

\paragraph{Base Retrievers}
We use a wide range of base retrievers in our experiments. We include a sparse retriever \textbf{BM25}, an unsupervised dense retriever \textbf{SPECTER-v2}~\cite{specter}, and the \textbf{Hybrid} of these two. We also include two instruction-tuned dense retrievers, \textbf{E5-large-v2}~\cite{E5} and \textbf{GritLM-7B}~\cite{GritLM}\footnote{We also experiment with other retrievers such as NV-Embed-v2~\cite{lee2025nvembed}. GritLM achieves the strongest performance on scientific paper retrieval, potentially because it specifically includes scientific corpora in its training data.}.

\paragraph{Baselines} We compare \Our with a collection of retrieval methods using \underline{corpus knowledge} and/or \underline{LLMs} to enhance base retrievers.
\begin{itemize}[leftmargin=*,nosep]
    \item \textbf{Boudin et al.}~\cite{boudin-etal-2020-keyphrase} uses a seq2seq \underline{keyphrase} generation model to enrich the corpus indexing.
    \item \textbf{BERT-QE}~\cite{zheng-etal-2020-bert} expands the query with \underline{relevant text chunks} selected from top-ranked papers returned by the base retriever.
    \item \textbf{ToTER}~\cite{toter} improves the retrieval performance with a \underline{topical taxonomy} by comparing the topic distributions of the query and documents predicted by a text classifier.
    \item \textbf{HyDE}~\cite{gao-etal-2023-precise} prompts an \underline{LLM} to generate hypothetical document that answers the query and encode it as the query vector.
    \item \textbf{GRF}~\cite{GRF} generates relevant context by an \underline{LLM}. We choose to generate scientific concepts for fair comparison.
    \item \textbf{CSQE}~\cite{lei-etal-2024-corpus} uses an \underline{LLM} to select relevant sentences from top-ranked documents of the base retriever, which are then used to expand the query together with a hypothetical document.
\end{itemize}

\paragraph{Evaluation Metrics} We use Recall@K (R@K) as our evaluation metric. Following previous studies, we use $K=50, 100$ for CSFCube and DORISMAE, and $K=5, 20, 100$ for LitSearch.

\paragraph{Implementation Details} For building semantic index, we train our text classifier using MAPLE~\cite{zhang2023effect} which contains topic labels from Microsoft Academic Graph~\cite{sinha2015overview}. We initialize it with \texttt{SPECTER-v2-base} and fine-tune it with learning rate at 5e-5 for 10 epochs. The balancing factor $\alpha$ is set to 1e-2. We also use \texttt{SPECTER-v2-base} for concept encoding. During retrieval, the number of candidate topics and key phrases is $k=50$. We use \texttt{GPT-4.1-mini} as the LLM for \Our and all LLM-based baselines for fair comparison. The experiments are run on one NVIDIA RTX A6000 when a GPU is needed.

\subsection{Retrieval Performance Comparison}

Table~\ref{table:baseline-res} shows the results of compared baselines with two retrievers, SPECTER-v2 and E5-large. We clearly see that \Our overall outperforms the compared baselines on all datasets. Specifically, we observe that methods using LLMs for query understanding achieves better performance, showing the strength of text understanding ability of LLMs in the retrieval task. Comparing with previous methods, \Our uses the LLM for concept-based query understanding and matching, which captures the query's need more explicitly and thus achieves stronger performance.

Table~\ref{table:base-retriever-res} additionally shows the results of \Our on different base retrievers. We can see that \Our consistently improves the performance of all kinds of retrievers. Even for GritLM-7B, the SOTA model reported in \citet{ajith-etal-2024-litsearch} and trained with scientific data, \Our still improves its performance on all datasets. Besides, \Our also improves the performance of Hybrid model, showing that concept-level semantic matching is not a combination of typical sparse and dense features, but in another intermediate level that really captures scientific knowledge.

\subsection{Ablation Studies}
We conduct ablation studies to show the effectiveness of each component of \Our. We include the following ablated versions:
\begin{itemize}[leftmargin=*,nosep]
    \item \textbf{No Topic}: excludes topics in semantic index.
    \item \textbf{No Phrase}: excludes phrases in semantic index.
    \item \textbf{No Corpus}: excludes the candidate scientific concepts from the corpus and thus prompts the LLM to directly generate scientific concepts based on its own knowledge. 
    \item \textbf{No LLM (class)}: excludes the LLM and selects scientific concepts for the query using the fine-tuned topic classifier.
    \item \textbf{No LLM (freq)}: excludes the LLM and uses the top-20 most frequent concepts mentioned by top-ranked papers returned by the base retriever.
\end{itemize}
Table~\ref{table:ablation-res} shows the results of ablation studies on the LitSearch dataset with SPECTER-v2 as the base retriever. First, we observe that either ablating topics or phrases from the semantic index will degrade the performance, with phrases affecting more because of its capturing more detailed information. Second, removing the augmented corpus knowledge from LLM prompting will greatly affect the final performance, because LLMs tend to generate terms not matched with the corpus. Finally, removing LLM but using topic classifier or statistic-based metric for query concept identification also drastically decreases the performance, showing the power of LLM on query understanding when augmented with corpus knowledge.

\begin{table}[!t]
\centering
\caption{Ablation studies of \Our on LitSearch.}
\label{table:ablation-res}
\vspace*{-0.5em}
\scalebox{0.8}{
\begin{tabular}{l|ccc}
\toprule
& R@5    & R@20    & R@100 \\
\midrule
SPECTER-v2 & 0.3931 & 0.5551 & 0.7205\\
\midrule
\bf Indexing\\
\quad No Topic & 0.4331 & 0.6040 & 0.7687\\
\quad No Phrase & 0.4160 & 0.5682 & 0.7447\\
\midrule
\bf Retrieval\\
\quad No Corpus & 0.4060 & 0.5557 & 0.7260\\
\quad No LLM (class) & 0.4079 & 0.5459 & 0.7320\\
\quad No LLM (freq) & 0.3897 & 0.5497 & 0.7267\\
\midrule
\Our & \bf 0.5028 & \bf 0.6316 & \bf 0.7746\\
\bottomrule
\end{tabular}
}
\vspace*{-1.25em}
\end{table}

\subsection{Efficiency Analysis} \label{sec:efficiency}

We show the efficiency of \Our by comparing it with other LLM-based baselines. Specifically, we report the following factors of each method: the number of base retriever calls per query (\textbf{\# RET}), the number of LLM calls per query (\textbf{\# LLM}), the average number of tokens generated by the LLM per query (\textbf{LLM Output Len}), and the average running time per query (\textbf{Running Time}). As stated in Sect.~\ref{sec:exp-setup}, we use the same LLM checkpoint for all baselines for a fair comparison. Table~\ref{table:efficiency} shows the detailed comparison results on LitSearch. We can clearly see that \Our takes the least inference time among the compared methods, with 1.5$\times$ faster than the second fast method. Not only does \Our call the retriever and LLM only once, it also expects minimal number of tokens responded by the LLM because of the concept-based structured input and output format.

Additionally, the offline indexing part of \Our is efficient as well. For semantic index construction on CSFCube, to get the candidate topics, the text classifier inference time is 52 seconds, and to prompt LLM for topic and keyphrase selection, it takes 84 seconds and \$1.74 dollars. In comparison, directly prompting LLMs to build such semantic index on CSFCube will take approximately 42 minutes and \$85 dollars.

\subsection{Combination with LLM-based Reranking}

Recent studies also use LLM for reranking retrieval results by prompting it to provide a new ranked list of top documents. To show that \Our can naturally integrate with such a method, we compare the performance of LLM-based reranking with and without \Our. Following~\cite{ajith-etal-2024-litsearch}, we provide top-100 papers to the LLM and use the prompt from \citet{sun-etal-2023-chatgpt}. Table~\ref{table:reranking-res} shows the results on LitSearch with 3 base retrievers. We can see \Our does not conflict with LLM-based reranking by consistently improving its performance. While typical reranking only improves recall within provided number of papers (R@100 unchanged), \Our can brings more relevant papers and also improves R@100.

\begin{table}[!t]
\centering
\caption{Efficiency analysis of LLM-based methods.}
\label{table:efficiency}
\vspace*{-0.5em}
\scalebox{0.7}{
\begin{tabular}{l|ccccccc}
\toprule
& \# RET & \# LLM & LLM Output Len & Running Time\\
\midrule
HyDE & \bf 1 & \bf 1 & 169.39 tok & \,\,4.02 sec\\
GRF & 2 & \bf 1 & \,\,79.72 tok & \,\,2.78 sec\\
CSQE & 2 & 2 & 462.21 tok & 11.27 sec\\
\Our & \bf 1 & \bf 1 & \bf \,\,\,18.92 tok & \bf \,\,\,1.82 sec\\
\bottomrule
\end{tabular}
}
\vspace*{-1.25em}
\end{table}

\subsection{Parameter Studies}
\begin{figure}[t]
\centering
    \includegraphics[width=0.49\linewidth]{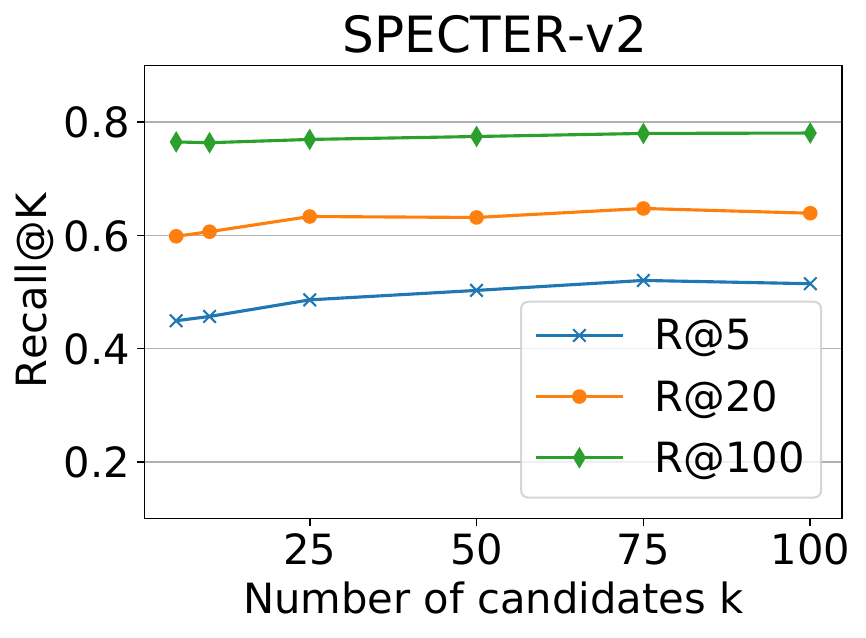}
    \hspace{-0.25em}
    \includegraphics[width=0.49\linewidth]{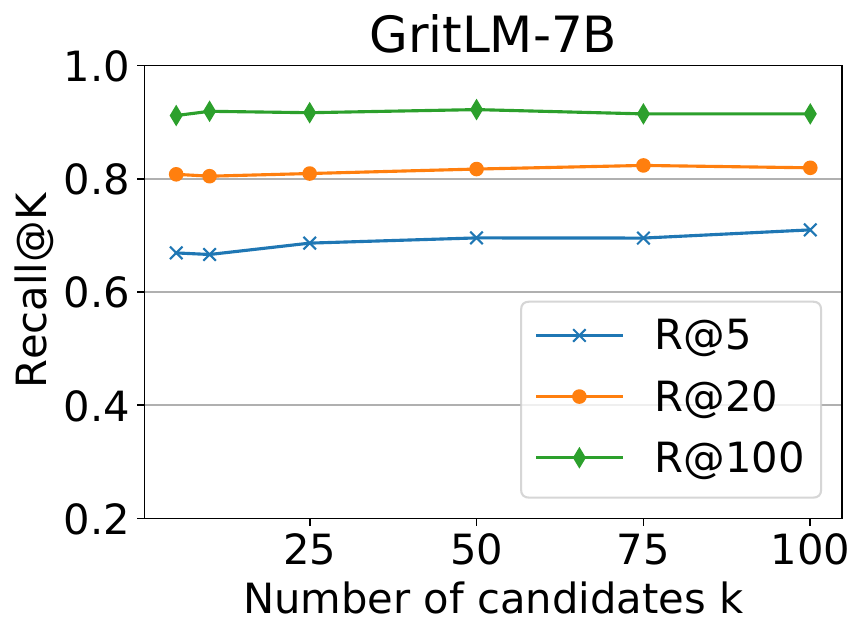}
    \hspace{-0.25em}
    \caption{Parameter analysis on LitSearch by varying $k$, the number of candidate query concepts.}
    \vspace{-1.25em}
    \label{fig:para_analysis}
\end{figure}

We study the influence of setting different number of candidates provided to LLM for query concept identification, \ie, $k$ in Sect.~\ref{sec:retrieval} for prompt in Figure~\ref{fig:prompt-core-concept}. We set the value of $k=5, 10, 25, 50, 75, 100$ and report their performance on LitSearch with base retriever SPECTER-v2 and GritLM. Results in Figure~\ref{fig:para_analysis} show that \Our overall is not very sensitive to the value of $k$ for $k \ge 25$. We notice slightly increased performance for smaller $k$ in the R@5 measures, which shows that an incomplete query concept set may not be sufficient to affect top-ranked documents that are hard to distinguish.
\section{Related Works}

\paragraph{Dense Retrieval}
Dense retrieval has become a core paradigm in modern information retrieval. Early models like Dense Passage Retrieval (DPR)~\cite{DPR} and ME-BERT~\cite{luan2021sparse} leveraged in-batch and BM25-based hard negatives to improve training efficiency. Subsequent methods refined negative sampling: ANCE~\cite{yu2021improving} used asynchronously updated indices; RocketQA~\cite{qu2020rocketqa} introduced cross-batch and denoised negatives; and ADORE~\cite{zhan2021optimizing} adopted dynamic sampling for greater stability. PAIR~\cite{ren2021pair} incorporated passage-level similarity signals, while FiD-KD~\cite{izacard2020distilling} distilled knowledge from reader models. 

In academic domains, specialized models like SciBERT~\cite{beltagy2019scibert} pre-trained on scientific texts laid the groundwork. 
\citet{parisot-zavrel-2022-multi} proposes a multi-objective approach that uses general-domain document relevance and scientific domain citation network and self-supervised data.
\citet{mandikal2024sparsemeetsdensehybrid} presents a hybrid approach that combines sparse and dense retrievers with a weighting parameter.
MixGR~\cite{cai-etal-2024-mixgr} matches queries and documents additionally at subquery and proposition levels and merge them with rank fusion.
Recent innovations also focus on knowledge distillation from cross-encoder rankers ~\cite{huang2024pairdistill, tao2022adam, zhang2023noisy}. Despite the progress and varied strategies for improving dense retrieval, these methods inherently rely on representing entire documents or queries with single dense embeddings, restraining them to capture fine-grained details crucial for accurately interpreting complex scientific queries. In contrast, our approach explicitly models the multi-granular scientific concepts within both queries and documents.

\begin{table}[!t]
\centering
\caption{Further analysis of \Our by combining with LLM-based reranking.}
\label{table:reranking-res}
\vspace*{-0.5em}
\scalebox{0.8}{
\begin{tabular}{l|ccc}
\toprule
& R@5    & R@20   & R@100\\
\midrule
SPECTER-v2 & 0.3931 & 0.5551 & 0.7205\\
\quad+ Reranking & 0.6636 & 0.7038 & 0.7205\\
\quad\quad+ \Our & \bf 0.6705 & \bf 0.7435 & \bf 0.7746\\
\midrule
E5-large-v2 & 0.5137 & 0.6573 & 0.7765\\
\quad+ Reranking & 0.6989 & 0.7480 & 0.7765\\
\quad\quad+ \Our & \bf 0.7108 & \bf 0.7963 & \bf 0.8312\\
\midrule
GritLM-7B & 0.6908 & 0.8001 & 0.9149\\
\quad+ Reranking & 0.7575 & 0.8470 & 0.9149\\
\quad\quad+ \Our & \bf 0.7774 & \bf 0.8520 & \bf 0.9221\\
\bottomrule
\end{tabular}
}
\vspace*{-1.25em}
\end{table}

\paragraph{LLM-Enhanced Retrieval}
Recent works show that LLMs can boost retrieval quality even when little or no human supervision is available. In zero-shot settings, HyDE~\cite{gao-etal-2023-precise} prompts an instruction-tuned LLM (e.g., InstructGPT) to imagine a hypothetical answer document, then encodes this synthetic text with an unsupervised dual-encoder; the dense representation guides nearest-neighbor search and already outperforms Contriever~\cite{Izacard2021UnsupervisedDI} without any task data.
At inference time, LLMs refine the query itself. Rewrite-Retrieve-Read pipeline~\cite{ma2023query} trains a lightweight rewriter with RL from the downstream reader LLM, consistently topping standard retrieve-then-read QA. For expansion, HyDE's hypothetical-document trick underpins GenRead~\cite{yu2022generate}, which sometimes matches or beats retrieval-based pipelines by generating context first. CSQE~\cite{lei-etal-2024-corpus} tempers hallucinations by mixing LLM expansions with sentences extracted from top-ranked corpus hits, outperforming fine-tuned neural expanders on tough TREC queries. CCQGen~\cite{kang2025improving} leverages an LLM to select topical concepts derived from a predefined top-down taxonomy. 
In our work, \Our leverages an LLM to select core concepts directly from candidate sets of both broad research topics and specific key phrases derived from the corpus, which augment LLMs with multi-granular knowledge to help query understanding and reduce hallucination.

\paragraph{Retrieval with Corpus Knowledge}
Recent work has explored leveraging corpus-based knowledge to enhance retrieval accuracy through query expansion and refinement techniques. Methods such as BERT-QE~\cite{zheng-etal-2020-bert} select corpus-contextualized text chunks to alleviate vocabulary mismatches, while GAR~\cite{mao2020generation} generates pseudo-passages for semantic enrichment of queries, demonstrating substantial gains. Query2doc~\cite{wang2023query2doc} further utilizes LLMs to create entire pseudo-documents, capturing external knowledge from web-scale training corpora. Similarly, classical pseudo-relevance feedback (PRF)~\cite{wang2021pseudo, lei-etal-2024-corpus} has been adapted to dense retrieval, reducing hallucinations and improving effectiveness. Graph-based methods such as \cite{macavaney2022adaptive, kulkarni2023lexically} utilize document similarity graphs to dynamically expand search results. 
ToTER~\cite{toter} designs taxonomy-based retrieval to identify the central topic classes and exploit their topical relatedness to supplement PLM-based retrievers. 
TaxoIndex~\cite{TaxoIndex} constructs a semantic index guided by an academic taxonomy, extracting and organizing concepts from documents, and then trains an indexing module to match these concepts with queries. 
In contrast, our proposed \Our framework avoids supervision, utilizing corpus-derived concepts in an unsupervised manner to build semantic index and perform pseudo-relevance feedback, thereby enhancing retrieval without additional training.

\section{Conclusion}

We present \Our, a novel scientific paper retrieval method that integrates LLM-guided query understanding with a concept-based semantic index. To overcome the limitations of existing methods, \Our identifies multi-granular scientific concepts to explicitly understand scientific queries at the concept level. By augmenting LLMs with corpus knowledge, \Our also facilitates LLM's understanding of query and context while reducing hallucination. Experiments demonstrate that \Our consistently improves retrieval performance across various base retrievers and outperforms various baseline methods while remaining highly efficient.

\section*{Limitations}
While \Our shows strong performance and efficiency in scientific paper retrieval, it also has several limitations. 
First, our current studies limit to scientific paper retrieval dataset with only title and abstract, while retrieving full scientific papers could be more challenging due to the difficulties of effectively understanding long structured text.
Second, \Our only considers scientific concepts as a set, while not considering their internal relationship which could bring more insights to paper and query understanding.
Third, although our use of LLMs is efficient, the reliance on prompting still introduces sensitivity to prompt design and model behavior, which may require tuning for different domains. 
Fourth, our experiments are done mainly on Computer Science domain corpora, because there is limited high-quality retrieval dataset available from other disciplines. We would like to argue that there is still a big gap in constructing paper retrieval benchmarks for different disciplines, which could be a research opportunity for future studies.
Finally, we only focus on English scientific paper retrieval in the work, while it remains a challenge on multilingual or multi-modal (e.g., figures, tables) paper retrieval.

\section*{Acknowledgments}
Research was supported in part by National Science Foundation IIS-19-56151, NSF IIS 25-37827, the Molecule Maker Lab Institute: An AI Research Institutes program supported by NSF under Award No. 2019897, and the Institute for Geospatial Understanding through an Integrative Discovery Environment (I-GUIDE) by NSF under BRIES Program No. HR0011-24-3-0325.  The research has used the Delta/DeltaAI advanced computing and data resource, supported  in part by the University of Illinois Urbana-Champaign and through allocation \#250851 from the Advanced Cyberinfrastructure Coordination Ecosystem: Services \& Support (ACCESS) program, which is supported by National Science Foundation grants OAC 2320345, \#2138259, \#2138286, \#2138307, \#2137603, and \#2138296.  Any opinions, findings, and conclusions or recommendations expressed herein are those of the authors and do not necessarily represent the views, either expressed or implied, of DARPA or the U.S. Government.

\bibliography{custom}

\appendix
\begin{figure*}[htbp]
\centering
\begin{tcolorbox}[
  enhanced,
  title=Query Core Concept Identification Prompt,
  attach boxed title to top center={yshift=-3mm,yshifttext=-1mm},
  colback=gray!15,
  colframe=gray!75!black,
  colbacktitle=gray!40,
  coltitle=black,
  fonttitle=\bfseries,
  boxed title style={size=small,colframe=gray!50!black},
  boxrule=0.5pt,
  arc=7pt,
  outer arc=7pt,
  left=10pt,
  right=10pt,
  top=10pt,
  bottom=10pt
]
You will receive a paper abstract along with a set of candidate topics for the paper. 

Your first task is to select the topics that best align with the core theme of the paper.
Exclude topics that are too broad or less relevant. 

Only use the topic names in the candidate set. 

Your second task is to generate a complete list of key phrases extracted from the paper.

Do some rationalization before outputting the list of relevant topics and key phrases.
\bigskip

Output format: `<top> topic 1, topic 2, ... </top>
<kp>key phrase 1, key phrase 2, ... </kp>'.
\bigskip

Paper: \hl{\{$d$\}}

\end{tcolorbox}
\caption{Prompts given to the LLM for building semantic index.}
\label{fig:prompt-sem-idx}
\end{figure*}

\begin{figure*}[htbp]
\centering
\begin{tcolorbox}[
  enhanced,
  title=Query Core Concept Identification Prompt,
  attach boxed title to top center={yshift=-3mm,yshifttext=-1mm},
  colback=gray!15,
  colframe=gray!75!black,
  colbacktitle=gray!40,
  coltitle=black,
  fonttitle=\bfseries,
  boxed title style={size=small,colframe=gray!50!black},
  boxrule=0.5pt,
  arc=7pt,
  outer arc=7pt,
  left=10pt,
  right=10pt,
  top=10pt,
  bottom=10pt
]
You will receive a query for research papers and a ranked list of papers returned by a retriever.

You will also be provided a list of research topics and key terms with their frequencies that are frequently mentioned by the top-ranked papers returned by the retriever.

Your task is to improve the provided retrieval results by selecting a list of topics and terms that can accurately identify the relevant papers of the query.

Make sure your selection is strictly based on the original query and does not contain repeated concepts.
\bigskip

Output format: `<ans>selection 1, selection 2, ...</ans>'.
\bigskip

Retriever result: \hl{\{$D^0(q)$\}}
\bigskip

Candidate topics: \hl{\{$T^0(q)$\}}
\bigskip

Candidate key terms: \hl{\{$P^0(q)$\}}
\bigskip

Original Query: \hl{\{$q$\}}

\end{tcolorbox}
\caption{Prompts given to the LLM for query core concept identification.}
\label{fig:prompt-core-concept}
\end{figure*}



\end{document}